\documentstyle[12pt]{article}
\makeatletter
\ifcase \@ptsize
\or % mods for 11 pt
\or % mods for 12 pt
\fi
\advance\textheight by \topskip

\ifcase \@ptsize
\or % mods for 11 pt
\or % mods for 12 pt
\fi

\def\be{\beta}
\def\al{\alpha}
\def\ep{\epsilon}
\def\la{\lambda}        
\def\de{\delta}         
\def\om{\omega}         
\def\sig{\sigma}

\def\an{$a_n^{(1)}$\ }
\def\d4{$d_4^{(1)}$\ }
\def\cn{$c_n^{(1)}$\ }

\def\noi{\noindent}
\begin{document}
\renewcommand{\thefootnote}{\fnsymbol{footnote}}
\baselineskip 20pt
\def\mb #1	{\mbox{\boldmath$#1$}}
\def\T		{^{^{\!\rm T\!}}}
\def\Insert#1	{\vspace{1in}\begin{center}\framebox{\bf #1 }\end{center}
\vspace{1in}}
\def\tbc	{\Insert{To Be Completed} }
\font\upright=cmu10 scaled\magstep1
\def\stroke{\vrule height8pt width0.4pt depth-0.1pt}
\def\topfleck{\vrule height8pt width0.5pt depth-5.9pt}
\def\botfleck{\vrule height2pt width0.5pt depth0.1pt}
\def\Ctext{{\rlap{\rlap{C}\kern 3.8pt\stroke}\phantom{C}}}
\def\Rtext{\hbox{\upright\rlap{I}\kern 1.7pt R}}
\def\Ntext{\hbox{\upright\rlap{I}\kern 1.7pt N}}
\def\oC{\ifmmode{{\hbox\Ctext}}\else\Ctext\fi}
\def\oN{\ifmmode{\vcenter{\Ntext}}\else\Ntext\fi}
\def\oR{\ifmmode{\vcenter{\Rtext}}\else\Rtext\fi}
\def\epsf#1 {\epsfxsize=\hsize\epsfbox{EPSF/#1 }}
{\pagestyle{empty}
\rightline {DTP-93-35}
\rightline {July 1993}
\vskip 0.9in
\centerline{\LARGE The Topological Charges of the \an}
\vskip 0.1in
\centerline{\LARGE Affine Toda Solitons.}
\vskip 0.9in
\centerline {\it \Large William A. McGhee\footnote{W.A.McGhee@durham.ac.uk}}
\centerline {\it Department of Mathematical Sciences,}
\centerline {\it University of Durham,}
\centerline {\it Durham DH1 3LE,}
\centerline {\it England.}
\medbreak
\vskip 1.0in
\centerline {\Large Abstract}
\medbreak
\noi The topological charges of the \an affine Toda solitons are
considered. A general formula is presented for the number of charges
associated with each soliton, as well as an expression for the charges
themselves. For each soliton the charges are found to lie in the corresponding
 fundamental representation, though in general these representations are not
filled. Each soliton's topological charges are invariant under cyclic
permutations of the simple roots plus the extended root or equivalently, under
 the action of
 the Coxeter element (with a particular ordering). Multisolitons are
considered and are found to have topological charges filling the remainder of
 the fundamental representations as well as the entire weight lattice. The
article concludes with a discussion of some of the other affine Toda theories.

\vfill
\newpage}

%
%%%%%%%%%%%%%%%%%%%%%%%%%%% SECTION 1 - INTRODUCTION %%%%%%%%%%%%%%%%%%%%%%%%%%
%
\section{Introduction.}

\bigbreak
\pagenumbering{arabic}
\noi Affine Toda field theory is one of the three Toda field theories
each distinguished by the underlying Lie algebra to which it is associated.
 All of the theories are integrable, though only two of the three are
conformal. Conformal Toda field theory, is a conformal theory associated with
a finite dimensional Lie algebra. The conformal invariance can be broken in
such a way as to preserve its integrability. The remaining theory is affine
Toda (AT) field theory which is associated with loop algebras. Finally, extra
fields can be added to the AT model to regain conformal invariance. The
resulting theory is known as Conformal Affine Toda (CAT) field theory. This
theory is associated with the Kac-Moody algebras.
\medbreak
\noi Solitons were first constructed by Hollowood \cite{holl1} using
Hirota's method for the \an series of AT models, with imaginary coupling.
In \cite{holl1}, not only were the soliton energy and momenta found to be
real despite the complex nature of the solutions, but the ratios of their
masses were calculated and shown to be equal
to the ratios of the fundamental Toda particles of the real coupling theory --
 that is those excitations obtained by expanding the potential term of the
theory's Lagrangian density about its minimum. The topological charges of the
 $a^{th}$ soliton were claimed to lie in the $a^{th}$ fundamental
representation though not, in general, to fill it. Aspects of quantisation of
the solitons were considered and taken up in later papers \cite{holl2,holl3}.
\medbreak
\noi As well as extending the methods of \cite{holl1} to the remaining
theories \cite{mmc}, other authors have approached the subject from differing
directions. Firstly, Hirota's method has been applied to the CAT model from
which the AT solitons can be derived \cite{sao1,sao2,sao3}. Secondly,
B\"acklund
transformations have been constructed for the \an AT model \cite{lot}.
Finally, following the methods of Leznov and Saveliev \cite{lezart} a general
solution to all of the Toda models associated with a simple Lie algebra has
been presented \cite{otu1}, and investigated \cite{otu2,ko,jwru2}.
\medbreak
\noi Except in \cite{holl1,nied}, and more recently \cite{jwru2} which proves
 a
conjecture appearing in \cite{otu2} relating to the charges, there has been
 little
information gathered on the topological charges of the solitons. The purpose
of this article, therefore, is to
obtain a better understanding of the topological charges of the solitons in
the simplest AT theory, that of \an.
\medbreak
\noi An upper bound for the number of charges of the $a^{th}$
soliton is found to be
$$\tilde{h}_a={h\over\gcd(a,h)}$$
where $h$ is the Coxeter number. Indeed, $\tilde{h}_a$ turns out to be equal
to the number of topological charges of the $a^{th}$ soliton. The origin of
this formula lies in the dependence of the analytic expression for the
soliton on the $a^{th}$ power of an $h^{th}$
root of unity. {}From studying the soliton solutions, the relationship between
the topological charges is deduced to be the map
$$\mbox{\boldmath$\tau$} : \al_j\rightarrow\al_{(j-1)\bmod h}\qquad
(0\leq j \leq h-1),
\eqno(1.0\mbox{a})$$
which is also an automorphism of the extended Dynkin diagram,
$\Delta(a_n^{(1)})$.
Therefore, for each soliton, once one topological charge is calculated the
rest immediately follow by application of (1.0a). The map {\boldmath$\tau$}
has the same
 effect as the action of the Coxeter element \cite{ped1} with the following
 ordering:
$$\om_{tc}=r_n r_{n-1} r_{n-2}\ldots r_3 r_2 r_1.$$
As a result, the topological charges lie in the same representation, which is
shown to be the $a^{th}$ fundamental representation.
The expression for the topological charges themselves is also derived, and
found
to be given by
$$t_a^{(k)}=\sum_{j=1}^{n}{a(h-j)\bmod h\over h}\thinspace\al_j-\sum_{l=1}
^{k-1}\sum_{j=1}^n
 \de_{a(h-j)\bmod h,\  h-l\gcd(a,h)}\  \al_j\eqno(1.0\mbox{b})$$
\noi where $k=1,\ldots,\tilde{h}_a.$ This allows calculation of charges to
 be carried out much more easily than through the use of {\boldmath$\tau$}.
\medbreak
\noi When there are widely separated solitons it is intuitive to expect
the total topological charge to be the sum of the topological charges of the
individual solitons. This statement is proved to be true. Using this result,
a double soliton composed of solitons whose topological charges fill up the
first and $n$-th fundamental representations can be constructed, which has
charges filling up the adjoint representation, and in particular
has $\{\pm\al_1,\pm\al_2,\ldots,\pm\al_n\}$
 as topological charges. Further combinations of solitons can therefore be
constructed which fill up the fundamental representations and the entire
weight lattice itself.
\medbreak
\noi The article closes with a short discussion of some of the other AT
theories.
%%%%%%%%%%%%%%%%%%%%%% SECTION 2 - AFFINE TODA SOLITONS  %%%%%%%%%%%%%%%%%%%%%
\vfill
\newpage
\section{Affine Toda Solitons.}
\bigbreak
\noi The Lagrangian density of affine Toda field theory can be written in
 the form

$${\cal L}={1\over2}(\partial_\mu \phi)\!\cdot\!(\partial^\mu\phi)-
{m^2\over\be ^2}
\sum_
  {j=0}^n n_j(e^{\be\al_j\cdot\phi}-1).$$

\noi The field $\phi(x,t)$ is an $n$-dimensional vector, $n$ being the
rank
of the finite Lie algebra $g$. The $\alpha_j$'s, for $j=1,...,n$ are
the simple roots of $g$;
$\alpha_0$ is chosen such that the inner products among the elements
of the set
\{$\al_0,\al_j$\} are described by one of the extended Dynkin
diagrams. It is expressible in terms of the other roots by the equation

$$\al_0=-\sum_{j=1}^n n_j\alpha_j\eqno(2.0\hbox{a})$$

\noi where the $n_j$'s are positive integers.
Both $\be$ and m are constants, $\be$ being the coupling constant.
\medbreak
\noi The inclusion of $\al_0$ distinguishes affine Toda field theory from
Toda field theory. Toda field theory is conformal and integrable, its
integrability implying the existence of a Lax pair, infinitely many
conserved quantities and exact solubility \cite{mop,wilson,ot}
(for further references see \cite{lezbook}).
The extended root
is chosen in such a way as to preserve the integrability of Toda field
theory (though not the conformal property), with the enlarged set of
roots $\{\al_0,\al_j\}$ forming an admissible root system \cite{mop}.
\medbreak
\noi Also, the conformal invariance can be recovered via the introduction
 of two additional fields $\eta$ and $\nu$. The resulting theory is once again
both conformally invariant and integrable. This is the CAT model
\cite{cat2,cat1}.
\medbreak
\noi When the coupling constant $\beta$ is replaced by $i\beta$, the
potential term,
$$V(\phi)\sim\sum_{j=0}^n n_j(e^{i\be\al_j\cdot\phi}-1)$$
\noi which in the real coupling case, upon considering real fields, is zero
only for $\phi=0$, now has
zeros for $\phi\in{2\pi i\over\beta}\Lambda^{*}_W$, ($\Lambda_W^{*}$ being the
co-weight lattice). The appearance of many minima of the potential is an
indication
that soliton solutions, interpolating from one minimum at $x=-\infty$ to
another at $x=+\infty$, may exist. The change in the field between
$x=\pm\infty$
is therefore proportional to an element of the co-weight lattice.
\medbreak
\noi With complex coupling, the equations of motion
$$\partial ^2 \phi - {im^2\over\beta}\sum_{j=0}^n n_j \alpha_j e^{i\beta
\alpha_j\cdot\phi}=0\eqno (2.0\hbox{b})$$
\noi can, under the following substitution for the field $\phi(x,t)$,
$$\phi=-{1\over i\be}\sum_{i=0}^n {2\over \alpha_i\cdot\alpha_i} \alpha_i
\ln\tau_i\eqno(2.0\hbox{c})$$
\noi be reduced to the following form:
$$\sum_{j=0}^n\alpha_j Q_j=0$$
\noi where
$$ Q_j = \left ({\eta_j\over \tau_j^2} (D_t^2-D_x^2)\tau_j\cdot
\tau_j-2m^2n_j \left (\prod_{k=0}^n \tau_k ^{-\eta_k\alpha_k\cdot\alpha_j}-1
\right )\right ).$$
\noi The operators $D_x$ and $D_t$ are introduced to ease calculation.
They are Hirota derivatives, defined by
$$D_x^m D_t^n f\cdot g=\left ({\partial \over \partial x}- {\partial
\over \partial x^{\prime}} \right ) ^m \left ({\partial \over \partial
t}- {\partial \over \partial t^{\prime}} \right ) ^n
f(x,t)g(x^{\prime},t^{\prime}) \left \vert \right . _{x=x^{\prime}
\atop t=t^{\prime}}.$$
\medbreak
\noi It is assumed \cite{holl1,mmc} that $Q_j=0 \ \forall j$,
although
this is not the
most general decoupling. (The existence of $n+1$ $\tau$-functions
(compared to the $n$-component field $\phi$) can be traced back to the $\nu$
field in the CAT model \cite{sao1}). The equations of motion can
now be reduced to the form,
$$\eta_j (D_t^2-D_x^2)\tau_j\cdot\tau_j-2m^2 n_j \left ( \prod_{k=0}^n
\tau_k^{-\eta_k\alpha_k\cdot\alpha_j}-1 \right )\tau_j^2=0.
\eqno(2.0\hbox{d})$$
\noi In the spirit of Hirota's method for finding soliton solutions
\cite{hir}, it is assumed that
$$\tau_j=1+\de_j^{(1)}e^\Phi \ep+\de_j^{(2)}e^{2\Phi} \ep^2+ .... +
\de_j^{(p_j)}e^{p_j\Phi} \ep^{p_j}\eqno(2.0\hbox{e})$$
\noi where $\Phi=\sig (x-vt)+\xi$ and $\de_j^{(k)} (1\leq k \leq p_j),
\sig , v$ and $\xi$ are arbitrary complex constants. The constant
$p_j$ is a positive integer and $\ep$ a dummy parameter. The
method employed is to solve (2.3) at successive orders in $\ep$.
\medbreak
\noi The parameters $\sigma, v$ amd $m$ are related by
$$\sigma^2 (1-v^2)=m^2 \lambda.$$
\noindent where $\lambda$ is an eigenvalue of $NC$. The matrices $N$ and $C$
are defined as

$\bullet \ N=\mbox{diag}(n_0,n_1,\ldots,n_n),$

$\bullet \ (C)_{ij}=\al_i\cdot\al_j.$

\medbreak
\noi This result is used in
showing that the ratios of the soliton masses are equal to the ratios of the
masses of the fundamental Toda particles, as the non-zero eigenvalues of $NC$
 were shown in
 \cite{bcds1,bcds} (for the A, D and E theories) to be the squared masses of
the
fundamental particles.
The requirement that $\tau_j$ be
bounded as $x\rightarrow\pm\infty$, requires $n_0\eta_j p_j = n_j\eta_0 p_0.$
\medbreak
\noi Finally, it is unnecessary to consider the solution corresponding to
$\la=0$, as it is always $\phi=0$.
\medbreak
\subsection{The \an\ solitons}
\bigbreak
The Dynkin diagram for \an\ is shown in Figure 1. The matrix $NC$ has
eigenvalues given by
$$\ \qquad\qquad\la_a=4\sin^2\left ({\pi a\over h}\right ),
\qquad\mbox{where }h=n+1\ \mbox{and}\ a=1,\ldots,h-1.$$
\bigbreak
\
\bigbreak
\centerline{
\begin{picture}(250,100)(0,-10)
\put ( 18,50){\line( 1, 0){24}}
\put ( 58,50){\line( 1, 0){24}}
\put ( 98,50){\line( 1, 0){14}}
\put (168,50){\line( 1, 0){24}}
\put (208,50){\line( 1, 0){24}}
\put (152,50){\line(-1, 0){14}}
\put (132,95){\line(5,-2){102}}
\put (118,95){\line(-5,-2){102}}
\put ( 10,50){\circle{10}}
\put ( 50,50){\circle{10}}
\put ( 90,50){\circle{10}}
\put (160,50){\circle{10}}
\put (200,50){\circle{10}}
\put (240,50){\circle{10}}
\put (125,98){\circle{10}}
\put (10,35){\makebox(0,0){$\alpha_1$}}
\put (50,35){\makebox(0,0){$\alpha_2$}}
\put (90,35){\makebox(0,0){$\alpha_3$}}
\put (160,35){\makebox(0,0){$\alpha_{n-2}$}}
\put (200,35){\makebox(0,0){$\alpha_{n-1}$}}
\put (240,35){\makebox(0,0){$\alpha_n$}}
\put (125,114){\makebox(0,0){$\alpha_0$}}
\put (125,50){\makebox(0,0){.....}}
\put (125,10) {\makebox(0,0){Figure 1: Affine Dynkin diagram for \an.}}
\end{picture}
}
\noi With $\eta_j=1 \ \forall j$, the tau equations of motion are
$$(D_t^2-D_x^2)\tau_j\cdot\tau_j=2m^2 (\tau_{j-1} \tau_{j+1}-\tau_j^2)$$

\noi {\em i.e.\ }those of \cite{holl1}. Using the expansion (2.4) with
$p_0=1$ for the
single soliton solutions, it is found that
$$\tau_j=1+\omega^j e^{\Phi}$$
\noindent where $\omega$ is an $h^{th}$ root of unity.
There are $n$ non-trivial solutions \cite{holl1} (equal to the number of
fundamental particles) with $\omega_a=\exp (2\pi ia/h)$
where
$1\leq a\leq n$.
These $n$ solutions to \an\ can be written in the form
$$\phi_{(a)}=-{1\over i\beta}\sum_{k=1}^r \alpha_j\ln\left ({1+w_a^j
e^{\Phi} \over 1+e^{\Phi}}\right ).$$
\noi The general $N$-soliton solution can be built up from the single
soliton solutions,
 having $\tau$-functions given by \cite{holl1},
$$\tau_j(x,t)=\sum_{\mu_1=0}^1\cdots\sum_{\mu_N}^1 \exp\left(\sum_{p=1}^N
\mu_p\om_p^j\Phi_p
 +\sum_{1\leq p<q\leq N}\mu_p\mu_q\ln A^{(pq)} \right)\eqno(2.1\mbox{a})$$
\noi where
$$A^{(pq)}=-{(\sig_p-\sig_q)^2-(\sig_p v_p-\sig_q v_q)^2-4m^2\sin^2{\pi\over
h}(a_p-a_q)
\over (\sig_p+\sig_q)^2-(\sig_p v_p+\sig_q v_q)^2-4m^2\sin^2{\pi\over h}
(a_p+a_q)}$$
\noi is the `interaction constant'. As well as non-static multisolitons
there exist also static configurations composed of different species (i.e.
$a_p\neq a_q$) of single solitons. The number of such configurations is given
by
$$\sum_{k=1}^n
\left(\begin{array}{c}
n \\ k
\end{array}
\right)
=2^n-1.$$

\noi For example, static double solitons take the form
$$\tau_j(x,t)=1+\omega_{a_1}^j e^{\Phi_1} +\omega_{a_2}^j e^{\Phi_2}+A^{(12)}
\omega_{a_1}^j\omega_{a_2}^j e^{\Phi_1+\Phi_2}.$$
\noi With both solitons having the same velocity, $\sigma_1\sqrt{\la_2}=
\sigma_2\sqrt{\la_1}$, and so
$$\Phi_2=\sqrt{\la_2\over\la_1}(\Phi_1-\xi_1)+\xi_2.\eqno(2.1\mbox{b})$$
\noi When the coefficient of $\Phi_1$ in (2.1b) is a positive integer
then these solutions arise directly from the Hirota method, that is when
$$\sin^2\left(a_1\pi\over h\right)=k^2\sin^2\left(a_2\pi\over h\right).$$
\noi If $a_2=h-a_1$, then
$$A^{(12)}=\cos^2\left(a_1\pi\over h\right)=1-{1\over 4}\la_{a_1}$$
\noi and after the shift $\Phi_1\rightarrow\Phi_1+\xi_1$,
$$\tau_j(x,t)=1+y_1\omega_{a_1}^je^{\Phi_1}+y_2\omega_{-a_1}^j e^{\Phi_1}+y_1
y_2 \left(1-{1\over 4}\la_{a_1}\right)e^{2\Phi_1}$$
\noi where $y_1=e^{\xi_1}$, and $y_2=e^{\xi_2}$. Notice that when $y_1$=0
 (or $y_2=0$), that is when the first (or second) soliton is sent off to
infinity, the above solution reduces to a that of a single soliton. These
`mass degenerate' solutions which appear for $h=2p$ and $h=6p$ \cite{sao2}
(another two such configurations appear in \cite{sao2} for \an, but are
similarly
derived by the method above) were discussed in \cite{sao2,sao3} when they
where viewed as different to those constructed in \cite{holl1}. In fact, as
shown above, they are special cases of more general static configurations
lying within the $N$-soliton solution (2.1a).
\bigbreak
\vfill

%
%%%%%%%%%%%%%%%%%%%%%%%%% TOPOLOGICAL CHARGES %%%%%%%%%%%%%%%%%%%%%%%%%%%%%%
%
\newpage
\section{Topological Charge.}

\bigbreak
\noi The topological charge of the solitons is defined by

$$t={\be\over 2\pi}\int_{-\infty}^{\infty} dx\partial_x\phi=
{\be\over 2\pi}(\lim_{x\rightarrow\infty}-\lim_{x\rightarrow-\infty})\phi(x,t)
,$$

\noi which, using the ansatz (2.0c), can be written in the following form:

$$t = -{1\over 2\pi i}\sum_{j=0}^n {2\over \al_j\cdot\al_j}(\lim_{x
\rightarrow\infty}-
\lim_{x\rightarrow-\infty})\ln \tau_j(x,t)\ \alpha_j.\eqno(3.1)$$

\subsection{The \an solitons.}

\noi In this subsection, the topological charges of the single solitons
will be
calculated. The method used is to find a relationship between the charges,
which is then used to deduce all the charges from just one -- the highest
charge (so called, as all charges are subsequently shown to be obtainable from
 the
highest charge by subtracting a sum of simple roots). An explicit expression
for
 the topological charges associated to each soliton is constructed, before the
 relationship between
 the charges is shown to be equivalent to the action of the Coxeter element
(with a particular ordering), so implying that the topological charges lie in
 the same
representation. This is shown for the $a^{th}$ soliton to be the $a^{th}$
fundamental representation.
 Lastly,
the topological charges of the multisolitons is considered.
\bigbreak

\subsubsection{Single soliton topological charge.}

\noi It will prove convenient in the subsequent discussion to write
equation (3.1) in the
form,
$$t=-{1\over 2\pi i}\sum_{j=1}^n (\lim_{x\rightarrow\infty}-\lim_{x\rightarrow
 -\infty})
\ (\ln |f_j(x,t)|+i\arg f_j(x,t))\ \al_j\eqno(3.1.1\mbox{a})$$
\noi where
$$|f_j(x,t)|={|\tau_j(x,t)|\over|\tau_0(x,t)|}\ \rightarrow\ 1\qquad\mbox{as}
\qquad
x\rightarrow
\pm\infty.$$
\noi The topological charge is therefore given by
$$t=-{1\over 2\pi}\sum_{j=1}^n(\lim_{x\rightarrow\infty}-\lim_{x\rightarrow-
\infty})\arg
 f_j(x,t)\ \al_j.$$
\noi In order to calculate the topological charges it is necessary to
understand the behaviour of the complex functions $f_j(x,t)$. At $t=0$
(assuming throughout that $\sig>0$),
 with $\xi=\xi_1+i\xi_2$, it is convenient to write
$$e^{\sig x +\xi_1 + i\xi_2}=ye^{i\xi_2}\ ,$$

\noi where $y\rightarrow 0$ as $x\rightarrow -\infty$, $y\rightarrow
\infty$ as $x\rightarrow \infty$,
and $\xi_2$ is chosen such that $-\pi<\xi_2\leq\pi$. It is also
convenient to write
$$\omega_a^j=e^{i\mu}\qquad\mbox{where }\mu={2\pi a j\over h}\bmod(2\pi).$$

\noi Before proceeding, it is worthwhile to make clear the idea behind
the calculation
 which follows. The function
$$f_j(x,t)={1+ye^{i(\mu+\xi_2)}\over 1+ye^{i\xi_2}}$$
has zeros whenever $\mu+\xi_2=\pi$ and $y=1$, and is undefined when $\xi_2=
\pi$ and $y=1$. In either case, $\phi(x,t)$ is undefined. The range of
$\xi$ can then be divided into sectors, the boundary of each sector being
the values of $\xi_2$ for which $f_j(x,t)$ is either zero or undefined. Now,
the topological charge is a measure of the change in the argument of
$f_j(x,t)$ as $x$ goes from $-\infty$ to $+\infty$, and so the
 topological charge can change only when the curve traced out by $f_j(x,t)$
in the complex
 plane, is either $(i)$ undefined, or $(ii)$ passes through the origin. The
implication therefore
is that the topological charge of the soliton is constant on each of the
sectors in the range of
$\xi_2$ mentioned above. Indeed, it will be shown that the topological charge
takes on a unique
 value in each sector. An expression for topological charge in one particular
sector, that of
the {\em highest charge}, is calculated and from it (in the following
subsection) an
expression for the remaining
charges is deduced.
\medbreak
\noi The number of sectors  $\tilde{h}_a$, which the range of $\xi_2$ is
 divided into is rather straightforward to calculate. It is equal to the
number of different values that
$${2\pi i a j\over h}\bmod 2\pi i$$
\noi can take, or in other words the smallest value of $q$ for which
$${2\pi i q a\over h} = 2\pi i k\qquad\hbox{where $q,k\in\oN$}.$$
Rewriting this as
$$q\tilde{a}=k\tilde{h}_a\ \ \mbox{where}\quad \tilde{a}={a\over\gcd(a,h)}
\quad
\mbox{and}\quad\tilde{h}_a={h\over\gcd(a,h)}$$
\noi are coprime, then $q=\tilde{h}_a$ and $k=\tilde{a}$. So the
 {\em maximum} number of values that the topological charge can take for the
$a^{th}$ soliton is
$$\tilde{h}_a={h\over\gcd(a,h)}.$$
The range of $\xi_2$ can therefore be split into $\tilde{h}_a$ regions
$$I_p=\left(-\pi+{2\pi p\over\tilde{h}_a},-\pi+{2\pi (p+1)\over\tilde{h}_a}
\right)$$\noindent with $0\leq p\leq \tilde{h}_a-1$. Consider now the
transformation
$$\xi_2\rightarrow\left(\xi_2+{2\pi a\over h}\right)\bmod(2\pi)\in[-\pi,\pi).
\eqno(3.1.1\hbox{b})$$
\noi If $\xi_2$ originally resides in the region $I_0$ then repeated
application of (3.1.1b) will send $\xi_2$ to each of the other regions in
turn, before returning to $I_0$ on the $\tilde{h}_a^{th}$ application. As
will now be shown, the above transformation is equivalent to a cyclic
permutation of the simple roots $\{\al_j\}$ plus the extended roots $\al_0$.
The $a^{th}$ soliton solution
takes the form
$$\phi_{(a)}(x,t)=-{1\over i\beta}\sum_{j=0}^n\al_j\ln(1+\omega_a^j
ye^{i\xi_2})$$
which, under the above transformation, becomes
$$
\phi_{(a)}(x,t) = -{1\over i\beta}\sum_{j=0}^n\al_j\ln(1+\omega_a^{j+1}
ye^{i\xi_2}) = -{1\over i\beta}\sum_{j=0}^n\al_{j-1}\ln(1+\omega_a^{j}
ye^{i\xi_2}),
$$
\noi the labelling of the roots being $\bmod h$. Therefore, to
calculate the full set of topological charges of a single soliton, all that
is required is to calculate the topological charge for one value of $\xi_2$
and then cyclically permute the labelling on the $\al_j\ (0\leq j\leq h-1)$
 to generate the
 others.
\medbreak
\noi Consider now the function $f_j(x,t)$. Splitting it up into its
 real and imaginary parts,
$$f_j(x,t)={1+y\thinspace [\thinspace\cos(\mu+\xi_2)+\cos\xi_2]+y^2\cos\mu
\over|1+ye^{i
\xi_2}|^2}+i\thinspace{y\thinspace [\thinspace\sin(\mu+\xi_2)-\sin\xi_2]+y^2
\sin\mu\over|1+
ye^{i\xi_2}|^2}.$$

\noi The imaginary part is zero for $y=0$ (i.e. at $x=-\infty$) and at
one other point given by
$$y=-\thinspace{\sin(\mu+\xi_2)-\sin(\xi_2)\over \sin\mu}\qquad(
\mbox{provided}\ y>0), \eqno(3.1.1\mbox{b})$$
\noi where $\mu\neq 0,\ \pi$. If $\xi_2=-\pi+\epsilon$, where $\epsilon>0$
 is an infinitesimal. Then,
\begin{eqnarray*}
\hbox{Im}(f_j(x,t))=0\qquad\mbox{for}\qquad y & = & 1-{(1-\cos\mu)\over\sin\mu}
\thinspace\epsilon
\thinspace  + \thinspace O(\epsilon^2)\ , \\
\mbox{and}\ \ \mbox{Re}(f_j(x,t))\thinspace |1+ye^{i\xi_2}|^2 & = & {2\epsilon
\over\sin\mu}(1-\cos\mu)\thinspace\thinspace +\thinspace O(\epsilon^2)\ .
\end{eqnarray*}
Therefore the complex function $f_j(x,t)$ crosses the real axis positively for
$0<\mu<\pi$ and negatively for $\pi<\mu<2\pi$.
Also, for small positive $y$,
$$\hbox{Im}(f_j(x,t))|1+ye^{i\xi_2}|^2=-y\sin\mu +(\mbox{higher order terms})$$
\noindent i.e. the function starts off with negative imaginary part for $0<\mu<
\pi$ and positive imaginary part for $\pi<\mu<2\pi$. Finally, if $\mu=0$ then
$f_j=1$ and contributes zero to the topological charge, whereas if $\mu=\pi$,
 the change in the function's argument is $+\pi$. Taking all of this
information together the topological charge in this sector is deduced to be
\begin{eqnarray*}
t_a^{(1)}& = & -{1\over 2\pi i}\sum_{j=0}^n \left({2\pi iaj\over h}\bmod 2\pi
 i\right)\ \al_j\\
         & = & \sum_{j=0}^n{a(h-j)\bmod h\over h}\al_j.
\end{eqnarray*}
\noi This topological charge will be called the `highest charge' since
the difference
 between it and all subsequent topological charges, is proportional to a sum
of positive
 roots.
The remaining charges are therefore generated under $\hbox{\boldmath$\tau$}:
\al_j\rightarrow
\al_{(j-1)\bmod h}$. The order of {\boldmath$\tau$} acting on the highest
charge is the
smallest value of $q$ such that
$$a(h-(j+q))\bmod h=a(h-j)\bmod{h}.$$
 This is given by $q=\tilde{h}_a$, confirming that $\tilde{h}_a$ is in fact
equal to the number of charges for the $a^{th}$ soliton.
\vfill
\newpage
\subsubsection{An explicit formula for the charges.}
\bigbreak
\noi Consider the highest charge, which is written for convenience in the
 following form:
$$\la_0\al_0+\la_1\al_1+\cdots+\la_n\al_n$$
\noi Each $\la_j$ is equal to one of $0,1/\tilde{h}_a,2/\tilde{h}_a,
\ldots,(\tilde{h}_a-1)/\tilde{h}_a$. The other $\tilde{h}_a-1$ charges are
obtained by cyclically permuting the labelling of the simple roots so that
$$\la_0=1/\tilde{h}_a,\ \la_0=2/\tilde{h}_a,\ldots,\ \la_0=(\tilde{h}_a-1)/
\tilde{h}_a.$$
\noi Consider now, the permutation that results in $\la_0=k/\tilde{h}_a$
 where $(1\leq k\leq \tilde{h}_a-1)$. This is in effect equivalent to adding,
 modulo $h$,
$${k\over \tilde{h}_a}(\al_0+\al_1+\ldots+\al_n)$$
\noi to the highest charge. Therefore,
$$\la_j\rightarrow\cases{\la_j+k/\tilde{h}_a, &if $\la_j+k/\tilde{h}_a <1$;
\cr\cr \la_j+k/\tilde{h}_a-1, &if $\la_j+k/\tilde{h}_a \geq1.$}\eqno{(3.1.2
\mbox{a})}$$

\noi Using (2.0a) to set $\la_0$ equal to zero, the overall effect of the
 permutation is the subtraction of $1$ from $\la_j$ where $\la_j+k/
\tilde{h}_a\geq 1$. The expression for the topological charges is therefore
 deduced to be

$$t_a^{(k)}=\sum_{j=1}^{n}{a(h-j)\bmod h\over h}\thinspace\al_j -
\sum_{l=1}^{k-1}\sum_{j=1}^n
 \de_{a(h-j)\bmod h,\  h-l\gcd(a,h)}\  \al_j$$
\noi where $k=1,\ldots,\tilde{h}$. The example of the $A_5$ theory is given
in Appendix A.1.

\subsubsection{The highest charge and its fundamental representation.}

\noi In this section it will be shown that the topological charge of the
$a^{th}$ soliton lies in the $a^{th}$ fundamental representation. This will
be used in the next section when the remaining solitons will be shown to lie
in the same representation as the highest charge and so imply that all the
topological charges lie in the same fundamental representation.
\medbreak
\noi It will be convenient to write $a=h-b$, $b=\tilde{b}\gcd(b,h)$, and
 $h=\tilde{h}
\gcd(b,h)$.

\noi Due to the symmetry of the \an theory under $\al_i\rightarrow
\al_{h-i}$ for
$1\leq i\leq n$, it is necessary only to consider $$b\leq\cases{{1\over 2}
(h-1), &if h is
 odd;\cr\cr {1\over 2}h, &if h is even.}\eqno{(3.1.2\mbox{a})}$$
The highest charge is then given by
$$t_a^{(1)}=\sum_{j=1}^n {bj\bmod h\over h}\al_j.$$
The inner products of $t^{(1)}$ with each of the simple roots will be
considered and shown to be transformable via Weyl reflections to the highest
weight of the $a^{th}$ fundamental representation. Consider firstly the case
of $\gcd(b,h)=1$, i.e. $b$ and $h$ coprime. Defining,
$$\Omega(k)=\left[{hk\over b}\right]$$
where $[...]$ denotes the integer part, then
$$t^{(1)}_a\cdot\al_j=\cases{
\ \  1 &\hbox{for $j=\Omega(k)$, \ \ \ \ \ where $k=1,\ldots,b-1$},\cr
\! -1 &\hbox{for $j=\Omega(k)+1$, where $k=1,\ldots,b-1$},\cr
\ \  1 &\hbox{for $j=h-1$},\cr
\ \  0 &\hbox{otherwise}.}$$
Also, $\Omega(k)+1<\Omega(k+1)$ for $k=1,\ldots,b-2$, and $\Omega(b-1)<h-1$.
\noi Therefore, in general, $t_a$ has inner products with the simple
roots of the form
$$t_a\cdot\{\al_j\}=(0,0,\ldots,0,1,-1,0,\ldots,0,1,-1,0,\ldots,0,0,1),
\eqno(3.1.2\hbox{b})$$
\noi the notation indicating that the $j^{th}$ component of the row
vector is given by $t_a\cdot\al_j$. There are two things that can be
immediately shown to be true. Notice that if a
weight, $w$ has inner products with the simple roots given by
$$w\cdot\{\al_j\}=(\ \ldots,0,1,-1,0,0,\ldots\ )$$
\noi then under a Weyl reflection in the root which has inner product
$-1$ with $w$,
$w\ \rightarrow\ w'$, where
$$w'\cdot\{\al_j\}=(\ \ldots,0,1,-1,0,\ldots\ ).$$

\noi Applying this to the case of $t_a$, then a series of Weyl
reflections will result
in $t_a\ \rightarrow\ t'_a$ where
$$t'_a\cdot\{\al_j\}=(0,\ldots,0,1,-1,1,-1,\ldots,1,-1,1).$$
\noi If a weight $w$ has inner product with the simple roots now given by
$$w\cdot\{\al_j\}=(\ldots,1,-1,1,\ldots)$$
\noi then again pereforming a Weyl reflection in the simple root which
has inner product
 $-1$ with $w$, $w\ \rightarrow\ w'$ where
$$w'\cdot\{\al_j\}=(\ldots,0,1,0,\ldots)$$
\noi This last procedure combined with the previous one, can be applied
to $t'_a$ to
finally give $t''_a$ which is expressed via
$$t''_a\cdot\{\al_j\}=(0,\ldots,0,1,0,\ldots,0)$$
\noi with the $1$ appearing in the $d^{th}$ position, $d$ being given by,
$$d=n-[b-1]=h-b=a.$$
\noi Therefore, $t_a$ lies in the same representation as $t''_a$, i.e.
the $a^{th}$
 fundamental representation.
\noi The generalisation to the case of $\gcd(b,h)\neq 1$ is
straightforward. If $\tilde{t}^{(1)}$ is the highest charge in the theory
with Coxeter number $\tilde{h}$ of the $\tilde{a}=\tilde{h}-\tilde{b}$
soliton, then the highest charge of the current soliton is given by
$$t^{(1)}=(\tilde{t}^{(1)},0,\tilde{t}^{(1)},0,\ldots,\tilde{t}^{(1)}).
\eqno(3.1.2\hbox{c})$$
Then by the results of the above discussion the inner products of $t^{(1)}$
 with the simple roots is also of the form (3.1.2b), and again by the above
 lies in the $a^{th}$ fundamental representation.
\bigbreak

\subsubsection{The Topological charges, the Coxeter element and the
fundamental \mbox{representations}.}
\bigbreak
\noi In the last two subsections, the topological charges of the $a^{th}$
 soliton were calculated, with the highest charge shown to lie in the $a^{th}$
 fundamental representation. In this subsection it will be shown that the
cyclical permutation of the roots used to connect the topological
charges is in fact equivalent to the application of the Coxeter element
$$\om_{tc}=r_n r_{n-1} r_{n-2}\ldots r_3 r_2 r_1.\eqno(3.1.4\mbox{b})$$
where $r_i$ is a Weyl reflection in the $i^{th}$ simple root $\al_i$. It is
important to note that the ordering of the Weyl reflections is not arbitrary
 -- other orderings do not necessarily connect the charges. This can be shown
 for the case of $A_5$ discussed in the previous subsection (see Appendix A.1
 for an illustration of this).
\medbreak
\noi Firstly, consider the effect of $\omega_{tc}$ on the set of simple
 roots $\{\al_j\}$.
 It can be shown
%\vskip -48pt
\begin{eqnarray*}
\al_0 & \rightarrow & \al_0+\al_1+\al_2+\ldots +\al_{n-1}+2\al_n,\\
\al_1 & \rightarrow & -\al_1-\al_2-\al_3-\ldots -\al_n, \\
\hbox{and}\qquad\al_i & \rightarrow & \al_{i-1}\ \ \mbox{for $2\leq i\leq n$,}
\end{eqnarray*}
and so an arbitrary linear combination of the simple roots plus the extended
 root
$$u=\la_0\thinspace\al_0+\la_1\al_1+\la_2\al_2+\ldots+\la_n\al_n$$ is
transformed thus:
\begin{eqnarray*}
u&& \!\!\!\!\rightarrow \la_0\al_0+(\la_0-\la_1+\la_2)\thinspace\al_1+\ldots
+(\la_0-\la_1+\la_n)\thinspace \al_{n-1}+(2\la_0-\la_1)\thinspace \al_n\\
&& \!\!\!\!= \la_1\al_0+\la_2\al_1+\la_3\al_2+\ldots +\la_n\al_{n-1}+\la_0\al_n
\end{eqnarray*}
by equation (2.0a). Using the notation
$$\la_0\al_0+\la_1\al_1+\la_2\al_2+\ldots+\la_n\al_n=(\la_0,\la_1,\la_2,
\ldots,\la_n),$$
then
$$\omega_{tc}(\la_0,\la_1,\la_2,\ldots\la_{n-1},\la_n)=(\la_1,\la_2,\la_3,
\ldots,\la_n,\la_0),$$
i.e. the action of the Coxeter element cyclically permutes the $\la_j$'s.
This invariance of the set of topological charge under the action of the
Coxeter element means that the topological charges lie in the same
representation as the highest charge i.e. the $a^{th}$ fundamental
representation.

\subsubsection{The other \an automorphisms}

\noi In subsection 3.1.1 it was shown that the set of topological charges
corresponding to each soliton was invariant under the automorphisms of the
extended
Dynkin diagram which cyclically permute the elements of the extended root
system. There
 are other
automorphisms of the extended diagram for \an. In this subsection, the
effect of
these mappings (up to a sign), on a soliton's topological charges will be
discussed.

\noi Sending $\xi_2\rightarrow-\xi_2$ in the soliton solution is equivalent
to evaluating the topological charge in region $I_{\tilde{h}_a-1-p}$ rather
than
$I_{p}$ where $0\leq p\leq\tilde{h}_a-1$. The form of the $a^{th}$ soliton
solution
is
$$\phi_{(a)}(x,t)=-{1\over i\beta}\sum_{j=1}^n\al_j\ln\left({1+\om_a^jye^{-i
\xi_2}\over
1+ye^{-i\xi_2}}\right)=-{1\over i\beta}\sum_{j=1}^n\al_j\ln\left(\om_a^j{1+
\om_a^{h-j}
y^{-1}e^{i\xi_2}\over 1+y^{-1}e^{i\xi_2}}\right).$$

\noi The last expression can be recast into the form
$$\phi_{(a)}(x,t)=-{1\over i\beta}\sum_{j=1}^n\al_{h-j}\ln\left(\om_a^{h-j}{1
+\om_a^jy^{-1}e^{i\xi_2}\over 1+y^{-1}e^{i\xi_2}}\right).$$
\noi The topological charge in region $I_{\tilde{h}_a-1-p}$ is therefore
obtained from the topological charge in region $I_p$ $(1\leq p\leq\tilde{h}_a-1
)$ via the mapping
$$\mbox{\boldmath$\sig_0$}:\al_j\rightarrow-\al_{h-j}.$$
\noi Combining this with the map of cyclic permutations of the extended root
system, \mbox{\boldmath{$\tau$}}, it is found that the set of topological
charges of each soliton is invariant under
$$\mbox{\boldmath$\sig_k$}:\al_j\rightarrow-\al_{(k-j)\bmod h}\qquad (0\leq
j,k\leq h-1).$$
\noi It is compelling to associate these mappings with the automorphisms of
the extended Dynkin diagrams which reflect the diagram in a line splitting it
in two as shown in figure 2, below.
\
\bigbreak
\centerline{
\begin{picture}(450,125)(0,-10)
\put ( 30,80){\line( 1, 1){14}}
\put ( 30,70){\line( 1,-1){14}}
\put ( 58,50){\line( 1, 0){24}}
\put ( 98,50){\line( 1, 0){14}}
\put (168,50){\line( 1, 0){24}}
\put (152,50){\line(-1, 0){14}}
\put (152,100){\line(-1, 0){14}}
\put ( 58,100){\line( 1, 0){24}}
\put ( 98,100){\line( 1, 0){14}}
\put (168,100){\line( 1, 0){24}}
\put (200,58){\line( 0, 1){34}}
\thicklines
\put ( 34,75){\line( 1, 0){78}}
\put (138,75){\line( 1, 0){59}}
\put (204,75){\line( 1, 0){10}}
\put ( 16,75){\line(-1, 0){10}}
\put (125,75){\makebox(0,0){.....}}
\put (10,75){\oval(20,20)[bl]}
\put (10,75){\oval(20,20)[tl]}
\put (10,85){\vector(1,0){6}}
\put (10,65){\vector(1,0){6}}
\thinlines
\put ( 25, 75){\circle{10}}
\put ( 50, 50){\circle{10}}
\put ( 90, 50){\circle{10}}
\put (160, 50){\circle{10}}
\put (200, 50){\circle{10}}
\put ( 50,100){\circle{10}}
\put ( 90,100){\circle{10}}
\put (160,100){\circle{10}}
\put (200,100){\circle{10}}
\put (25,60){\makebox(0,0){$\al_k$}}
\put (125,50){\makebox(0,0){.....}}
\put (125,100){\makebox(0,0){.....}}
\put (125,15) {\makebox(0,0){\an when $n$ is even.}}
\put (280,80){\line( 1, 1){14}}
\put (280,70){\line( 1,-1){14}}
\put (430,80){\line(-1, 1){14}}
\put (430,70){\line(-1,-1){14}}
\put (308,50){\line( 1, 0){24}}
\put (348,50){\line( 1, 0){14}}
\put (402,50){\line(-1, 0){14}}
\put (402,100){\line(-1, 0){14}}
\put (308,100){\line( 1, 0){24}}
\put (348,100){\line( 1, 0){14}}
\thicklines
\put (284,75){\line( 1, 0){78}}
\put (388,75){\line( 1, 0){36}}
\put (444,75){\line( 1, 0){10}}
\put (266,75){\line(-1, 0){10}}
\put (375,75){\makebox(0,0){.....}}
\put (260,75){\oval(20,20)[bl]}
\put (260,75){\oval(20,20)[tl]}
\put (260,85){\vector(1,0){6}}
\put (260,65){\vector(1,0){6}}
\thinlines
\put (275, 75){\circle{10}}
\put (300, 50){\circle{10}}
\put (340, 50){\circle{10}}
\put (410, 50){\circle{10}}
\put (435, 75){\circle{10}}
\put (300,100){\circle{10}}
\put (340,100){\circle{10}}
\put (410,100){\circle{10}}
\put (275,60){\makebox(0,0){$\al_k$}}
\put (375,50){\makebox(0,0){.....}}
\put (375,100){\makebox(0,0){.....}}
\put (342,10) {\makebox(0,0){\an when $n$ is odd.}}
\put (215,-10){\makebox(0,0){Figure 2: Reflection symmetry of the \an Dynkin
 diagram.}}
\end{picture}
}

\bigbreak
\subsubsection{Multisoliton solutions.}
\bigbreak
\noi In this section, a multisoliton configuration composed of $N$
 widely separated solitons is considered. In this large separation
approximation the topological charge of the configuration as a whole is found
 to be the sum of the topological charges of the individual solitons. This
will be done via an inductive arguement. In \cite{holl1} the tau functions of
 the multisolitons were found to be
$$\tau_j(x,t)=\sum_{\mu_1=0}^1\cdots\sum_{\mu_N}^1 \exp\left(\sum_{p=1}^N
\mu_p\om_p^j\Phi_p +\sum_{1\leq p<q\leq N}\mu_p\mu_q\ln A^{(pq)} \right)$$
\noi where
$$A^{(pq)}=-{(\sig_p-\sig_q)^2-(\sig_p v_p-\sig_q v_q)^2-4m^2\sin^2{\pi\over
n+1}(a_p-a_q)
\over (\sig_p+\sig_q)^2-(\sig_p v_p+\sig_q v_q)^2-4m^2\sin^2{\pi\over n+1}
(a_p+a_q)}$$
\noi is the `interaction constant'. Relabelling the solitons, if necessary,
then
$$\sigma_1 v_1 <\sigma_2 v_2 <\ldots <\sigma_{N-1} v_{N-1} <\sigma_{N} v_{N}.
\eqno(3.1.5\hbox{a})$$
\noi It will be convenient to write $e^{\sigma_i(x-v_i t +\xi^{(i)})}=
ye^{-\mu_i(t)}e^{i\xi_2^{(i)}}$, where $\mu_i(t)=\sigma_i v_i t -
\xi_1^{(i)}$. If $t=T$ is fixed for sufficiently large $T$, then write
$\mu_i(T)=\mu_i$ so that
$$\mu_1<\!\! <\mu_2<\!\! <\ldots <\!\! <\mu_{N-1}<\!\! <\mu_{N}.\eqno(3.1.5
\hbox{b})$$
\noi It is worthwhile to find the range of $y$ for which the soliton
field $\phi(x,t)$ has its most rapid variation (and so where the soliton is
located). This is done via the parameter $k>\!\!>1$, and the imposition that
$${1\over k}<|\omega_{a_i}^j y^{\sig_i} e^{-\mu_i} e^{i\xi_2^{(i)}}|<k,$$
since below the lower limit ${\tau_j/\tau_0}\sim 1$, and above the upper
limit $\tau_j/\tau_0\sim\om_{a_i}^j$. The corresponding limits in the range
of $y$ are:
\begin{eqnarray*}
y\sim\left( {1\over k}e^{\mu_i}\right )^{1/\sig_i} &&\hbox{for}\qquad|
\omega_{a_i}^j y^{\sigma_i}e^{-\mu_i}e^{i\xi_2^{(i)}}|\sim {1\over k},\\
\hbox{and}\qquad y\sim\left( {k}e^{\mu_i}\right )^{1/\sig_i} &&\hbox{for}
\qquad |\om_{a_i}^j y^{\sigma_i}e^{-\mu_i}e^{i\xi^{(i)}}|\sim {k}.
\end{eqnarray*}\noindent The point in time considered $T$, can be chosen large
 enough so that each of the above regions are far apart, that is
\begin{eqnarray*}
\left({1\over k}e^{\mu_1}\right)^{1/\sig_1}<(ke^{\mu_1})^{1/\sig_1}<\!\!<\left
({1\over k}e^{\mu_2}\right)^{1/\sig_2}&&\!\!\!\!\!\!\!\!\!\!<(ke^{\mu_2})^{1/
\sig_2}<\!\!<\ldots\\
\ldots\left({1\over k}e^{\mu_{N-1}}\right)^{1/\sig_{N-1}}&&\!\!\!\!\!\!\!\!\!
\!<(ke^{\mu_{N-1}})^{1/\sig_{N-1}}<\!\!<\left({1\over k}e^{\mu_{N}}
\right)^{1/\sig_{N}}<(ke^{\mu_{N}})^{1/\sig_{N}}.
\end{eqnarray*}
\noi The scene is now set for a straightforward calculation of the
multisoliton topological charge. Consider the two soliton solution
$${\tau_j\over\tau_0}={{1+\omega_{a_1}^j y^{\sigma_1} e^{-\mu_1}
e^{i\xi_2^{(1)}}
+\omega_{a_2}^j y^{\sigma_2} e^{-\mu_2} e^{i\xi_2^{(2)}}(1+A_{12}
\omega_{a_1}^j
y^{\sigma_1} e^{-\mu_1} e^{i\xi^{(1)}_2})}\over{1+y^{\sigma_1} e^{-\mu_1}
e^{i\xi_2^{(1)}}+ y^{\sigma_2} e^{-\mu_2} e^{i\xi_2^{(2)}}(1+A_{12}
y^{\sigma_1}
e^{-\mu_1} e^{i\xi_2^{(1)}})}}.$$
Here the first soliton is located in the range $\left({1\over k} e^{\mu_1}
\right)^{1/\sigma_1}\leq y \leq (ke^{\mu_1})^{1/\sig_1},$ and the second in
the range $\left({1\over k} e^{\mu_1}\right)^{1/\sigma_1}\leq y \leq
(ke^{\mu_1})^{1/\sig_1}$. Outside these regions $\tau_j/\tau_0$ is effectively
 constant and equal to (in order of increasing $y$), $1,\ \om_{a_1}^j,
\hbox{and}\  \om_{a_1}^j\om_{a_2}^j$, respectively. In the range
$\left({1\over k} e^{\mu_1}\right)^{1/\sigma_1}\leq y \leq (ke^{\mu_1})^{1/
\sig_1},$
$${\tau_j\over\tau_0}\sim{{1+\omega_{a_1}^j y^{\sig_1}e^{-\mu_1}
e^{i\xi_2^{(1)}}}
\over{1+y^{\sig_1}e^{-\mu_1}e^{i\xi_2^{(1)}}}}$$
\noindent (i.e. it is effectively the $j^{th}$ component of the first soliton)
 which contributes to the topological charge by $t_1$.
\noi Finally, for $\left({1\over k} e^{\mu_2}\right)^{1/\sigma_2}\leq y
\leq (ke^{\mu_2})^{1/\sig_2},$
$${\tau_j\over\tau_0}\sim\omega_{a_1}^j {{1+A_{12}\omega_{a_2}^j y^{\sig_2}
e^{-\mu_2}e^{i\xi_2^{(2)}}}\over{1+A_{12} y^{\sig_2}e^{-\mu_2}
e^{i\xi_2^{(2)}}}}$$
\noi which contributes $t_2$ to the topological charge. Therefore the
topological charge of the double soliton is given by
$$t=t_1+t_2.$$
\noi Suppose now that the $(N-1)$-soliton solution has topological charge
 given by
$$t_{(N-1)}=t_1+t_2+\ldots+t_{(N-1)}.$$
\noi If $\tau_j^{(p)}$ is the $j^{th}$ tau function of the $p$-soliton
solution, then for $0<y\leq (ke^{\mu_{N-1}})^{1/\sig_{N-1}}$,
$${\tau_j^{(N)}\over \tau_0^{(N)}}\sim {\tau_j^{(N-1)}\over \tau_0^{(N-1)}},$$
\noi contributing $t_{(N-1)}$ to the topological charge.
\medbreak
\noi
In the regions \mbox{$(ke^{\mu_{N-1}})^{1/\sig_{N-1}}<y<\left({1\over k}e^{
\mu_N}\right)^{1/\sig_N}\!\!,$} and for $y>(ke^{\mu_N})^{1/\sig_N}$,
the functions $\tau_j^{(N)}/\tau_0^{(N)}$ are effectively constant and equal
to \ $\omega_{a_1}^j\omega_{a_2}^j\ldots\omega_{a_{N-1}}^j$\  and \
$\omega_{a_1}^j\omega_{a_2}^j\ldots\omega_{a_{N-1}}^j\omega_{a_{N}}^j$
respectively. For $\left({1\over k}e^{\mu_{N}}\right)^{1/\sig_N}\leq y\leq
(ke^{\mu_N})^{1/\sig_N}$,
$${\tau_j^{(N)}\over\tau_0^{(N)}}\sim\omega_{a_1}^j\omega_{a_2}^j\ldots
\omega_{a_{N-1}}^j{{1+A_{a_1 a_2} \ldots A_{a_N a_{N-1}}\omega_{a_N}^j
y^{\sig_N}e^{-\mu_N}e^{i\xi_2^{(N)}}}\over{1+A_{a_1 a_2} \ldots A_{a_N a_{N-1}}
 y^{\sig_N}e^{-\mu_N}e^{i\xi_2^{(N)}}}}$$
\noi which contributes $t_N$ to the topological charge. Therefore the
topological charge of the $N$-soliton solution is
$$t=t_1+t_2+\ldots+t_{N-1}+t_N.$$
\noi This result still holds if the strict inequalities of (3.1.5a) are
 relaxed to allow for solitons to have $\sigma_i v_i$ equal, provided
$\xi_1^{(i)}$ is large enough so that (3.1.5b) holds.

\subsubsection{Multisolitons and representation theory.}

\noi Having established in the previous subsection that the topological
charge of a
multisoliton is the sum of the topological charges of its constituent solitons,
 the
representations in which the topological charges of these solitons lie will be
 discussed.
\medbreak
\noi Denote the set of topological charges of the $N$-soliton solution,
which is
composed of the $\{a_1,a_2,\ldots,a_N\}$-solitons, and the $a^{th}$ fundamental
 representation
 by
$${\cal{T}}_{(a_1,\ldots,a_N)}\qquad\mbox{and}\qquad{\cal{R}}_a$$
\noi respectively. As the topological charges of two widely separated solitons
 are equal to the pairwise sums of the topological charges of the individual
solitons, then for the special case of a double soliton composed of the single
 solitons associated to the first and $n$-th fundamental representations, the
resulting topological charges are the weights of the tensor product
representation ${\cal R}_1\otimes{\cal R}_n$. However, this tensor product of
 representations contains the adjoint representation, and so contains
$\{\pm\al_1,\pm\al_2,\ldots,\pm\al_n\}$. As a result, further
\mbox{multisoliton} configurations can be constructed that employ these
solitons having charges equal to the simple roots, and so fill up all the
fundamental representations as well as the entire weight lattice.
\bigbreak
\section{Discussion and conclusions.}

\medbreak
\noi Having discussed the \an theory in detail it is worthwhile to
consider what can be deduced about the other theories. Firstly consider
 \cn whose solitons are multisolitons of the $a^{(1)}_{2n-1}$ theory. As a
result
 it may be expected that the results of $\S 3.1.5$ can be used to calculated
the topological charges. This is not true, as the results on the multisolitons
 depend on the individual solitons being widely spaced -- the \cn single
solitons are $a^{(1)}_{2n-1}$ double solitons of {\em zero} separation. Indeed
it
can be shown that the number of topological charges of the $a^{th}$ soliton
$(1\leq a\leq n-1)$ in the \cn theory is given by
$$2(\tilde{n}-\tilde{a}+1)(1-{1\over 2}\delta_{0,\tilde{a}\bmod 2})$$
\noi where
$$\tilde{n}={n\over\gcd (a,n)}\qquad\hbox{and}\qquad\tilde{a}={a\over\gcd (a,n)
}$$

\noi and two topological charges for the $n^{th}$ soliton. Therefore, in
general, the number of topological charges in the \cn
theory is less than that calculated naively from the \an theory. The set
of topological charges corresponding to each soliton of the \cn theory is also
invariant under mappings related to the automorphisms of the theory. However,
there are not enough mappings to allow all of the charges to be deduced from
just one.
\medbreak
\noi For the $d_4^{(1)}$ and $d_5^{(1)}$ theories, the topological charges of
the
single solitons have been calculated and can be shown to lie in the
corresponding fundamental representation (for example see \cite{holl1} for
$d^{(1)}_4$) though, as in \an,
 not fill it. It is possible, as in the $a^{(1)}_n$ case to deduce
relationships
between the charges from the soliton solution, and so allow deduction of all
the charges from just one. There also exist products of reflections whose
orbits are made up of the charges, though the results are not as neat as those
 for \an.
\medbreak
\noi{\bf\large\bf Acknowledgements}
\medbreak
\noi I would like to thank the Science and Engineering Research Council
 for a studentship as well as Ed Corrigan, Patrick Dorey, G\'{e}rard Watts and
Michael Young for helpful comments and suggestions. Finally, my thanks go to
 Catherine Hemingway for her constant encouragement.

\vfill
\newpage
\noi{\large\bf Appendix}
\medbreak
\noi{\bf A.1 The topological charges of
{\boldmath$a_{\mbox{\bf 5}}^{\bf (1)}$} single solitons.}

\noi Consider the case of the $a_5^{(1)}$ theory whose Dynkin diagram is
shown in
figure
2, below.

\centerline{
\begin{picture}(180,100)(0,-10)
\put ( 18,50){\line( 1, 0){24}}
\put ( 58,50){\line( 1, 0){24}}
\put ( 98,50){\line( 1, 0){24}}
\put (138,50){\line( 1, 0){24}}
\put ( 10,50){\circle{10}}
\put ( 50,50){\circle{10}}
\put ( 90,50){\circle{10}}
\put (130,50){\circle{10}}
\put (170,50){\circle{10}}
\put (10,35){\makebox(0,0){$\alpha_1$}}
\put (90,35){\makebox(0,0){$\alpha_3$}}
\put (50,35){\makebox(0,0){$\alpha_2$}}
\put (130,35){\makebox(0,0){$\alpha_4$}}
\put (170,35){\makebox(0,0){$\alpha_5$}}
\put (10,65){\makebox(0,0){6}}
\put (50,65){\makebox(0,0){3}}
\put (90,65){\makebox(0,0){2}}
\put (130,65){\makebox(0,0){3}}
\put (170,65){\makebox(0,0){6}}
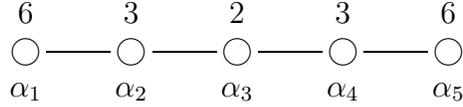
\put (90,10) {\makebox(0,0){Figure 2 : Dynkin diagram for $a_5^{(1)}$}}
\end{picture}
}
\noindent The number above each spot on the diagram is the number of
topological charges corresponding to that soliton.

\begin{tabbing}
\noindent\= First soliton (a=1).\qquad\qquad\qquad\qquad\qquad\qquad\qquad
\=\= Second soliton (a=2).\\
\>$t^{(1)}=\ \ \ $\=${5\over 6}\al_1+{2\over 3}\al_2+{1\over 2}\al_3+{1\over 3}
\al_4+{1\over 6}\al_5$,\> $t^{(1)}=\ \ \ $\=${2\over 3}\al_1+{1\over 3}\al_2
+{2\over 3}\al_4+{1\over 3}\al_5$,\\
\>$t^{(2)}=-$\>${1\over 6}\al_1+{2\over 3}\al_2+{1\over 2}\al_3+{1\over 3}\al_4
+{1\over 6}\al_5$,\> $t^{(2)}=-$\>${1\over 3}\al_1+{1\over 3}\al_2-{1\over 3}
\al_4+{1\over 3}\al_5$,\\
\>$t^{(3)}=-$\>${1\over 6}\al_1-{1\over 3}\al_2+{1\over 2}\al_3+{1\over 3}\al_4
+{1\over 6}\al_5$,\> $t^{(3)}=-$\>${1\over 3}\al_1-{2\over 3}\al_2-{1\over 3}
\al_4-{2\over 3}\al_5$.\\
\>$t^{(4)}=-$\>${1\over 6}\al_1-{2\over 6}\al_2-{3\over 6}\al_3+{2\over 6}\al_4
+{1\over 6}\al_5$,\> Third soliton (a=3).\\
\>$t^{(5)}=-$\>${1\over 6}\al_1-{2\over 6}\al_2-{3\over 6}\al_3-{4\over 6}\al_4
+{1\over 6}\al_5$,\> $t^{(1)}=$\>${1\over 2}\al_1+{1\over 2}\al_3+{1\over 2}
\al_5$,\\
\>$t^{(6)}=-$\>${1\over 6}\al_1-{2\over 6}\al_2-{3\over 6}\al_3-{4\over 6}\al_4
-{5\over 6}\al_5$.\> $t^{(2)}=-$\>${1\over 2}\al_1-{1\over 2}\al_3-{1\over 2}
\al_5$.
\end{tabbing}
\noindent The topological charges of the fourth and fifth solitons are
obtained from those of the second and first solitons respectively, via
$\al_j\rightarrow \al_{h-j}\ (1\leq j\leq n)$.
\bigbreak

\noindent{\bf A.2 The action of the Coxeter element.}

\noindent It is shown in section 3.1.4 that the action of the Coxeter element
with the ordering of (3.1.4a) generates all of the topological charges from
just one. Consider now the action of the Coxeter element
$$w=r_4r_2r_5r_3r_1$$
\noindent on the Weyl orbit of the second fundamental weight $\la_{(2)}$ of
the $a_5^{(1)}$ theory (this
 ordering is the familiar `black-white' ordering of \cite{ped1} i.e. the sets
of simple roots $\{\al_1,\al_3,\al_5\}$ and $\{\al_2,\al_4\}$ corresponding to
 $\{r_1,r_3,r_5\}$ and $\{r_2,r_4\}$ are composed of elements which are
orthogonal to each other). The Weyl orbit is partitioned into three Coxeter
orbits, say $C_+$, $C_-$, and $C_0$. This is visualised in Figure 3 where

$\bullet$ each spot corresponds to a weight in the Weyl orbit of $\la_{(2)}$,

$\bullet$ if two spots are joined by a line, they are Weyl reflections of each
 other in the simple \hbox{\ }\ \ \ \ \ \  root $\al_j$ where $j$ is the number
 on the line,

$\bullet$ spots with the same labelling (either $+$, $-$, or $0$) lie in the
same Coxeter orbit.
\medbreak
\noindent Therefore, whereas $t^{(1)}$ and $t^{(3)}$ lie in the same Coxeter
orbit, $t^{(2)}$ lies in a different one.
\centerline{
\begin{picture}(180,200)(40,-10)
\put ( 13,147){\line( 1,-1){24}}
\put ( 73,147){\line( 1,-1){24}}
\put (133,147){\line( 1,-1){24}}
\put (193,147){\line( 1,-1){24}}
\put ( 43,117){\line( 1,-1){24}}
\put (103,117){\line( 1,-1){24}}
\put (163,117){\line( 1,-1){24}}
\put ( 43,123){\line( 1, 1){24}}
\put (103,123){\line( 1, 1){24}}
\put (163,123){\line( 1, 1){24}}
\put (223,123){\line( 1, 1){24}}
\put ( 73, 87){\line( 1,-1){24}}
\put (133, 87){\line( 1,-1){24}}
\put ( 73, 93){\line( 1, 1){24}}
\put (133, 93){\line( 1, 1){24}}
\put (193, 93){\line( 1, 1){24}}
\put (103, 57){\line( 1,-1){24}}
\put (103, 63){\line( 1, 1){24}}
\put (163, 63){\line( 1, 1){24}}
\put (133, 33){\line( 1, 1){24}}
\put ( 10,150){\circle{8}}
\put ( 40,120){\circle{8}}
\put ( 70,150){\circle{8}}
\put ( 70, 90){\circle{8}}
\put (100,120){\circle{8}}
\put (100, 60){\circle{8}}
\put (130,150){\circle{8}}
\put (130, 90){\circle{8}}
\put (130, 30){\circle{8}}
\put (160,120){\circle{8}}
\put (160, 60){\circle{8}}
\put (190,150){\circle{8}}
\put (190, 90){\circle{8}}
\put (220,120){\circle{8}}
\put (250,150){\circle{8}}
\put (265,150){\makebox(0,0){$\la_{(2)}$}}
\put ( 10,150){\makebox(0,0){$\scriptscriptstyle +$}}
\put ( 40,120){\makebox(0,0){$\scriptscriptstyle +$}}
\put ( 70,150){\makebox(0,0){$\scriptscriptstyle -$}}
\put ( 70, 90){\makebox(0,0){$\scriptscriptstyle -$}}
\put (100,120){\makebox(0,0){$\scriptscriptstyle +$}}
\put (100, 60){\makebox(0,0){$\scriptscriptstyle -$}}
\put (130,150){\makebox(0,0){$\scriptscriptstyle 0$}}
\put (130, 90){\makebox(0,0){$\scriptscriptstyle 0$}}
\put (130, 30){\makebox(0,0){$\scriptscriptstyle 0$}}
\put ( 85, 90){\makebox(0,0){$t^{\scriptscriptstyle (3)}$}}
\put (145, 90){\makebox(0,0){$t^{\scriptscriptstyle (2)}$}}
\put (205, 90){\makebox(0,0){$t^{\scriptscriptstyle(1)}$}}
\put (160,120){\makebox(0,0){$\scriptscriptstyle +$}}
\put (160, 60){\makebox(0,0){$\scriptscriptstyle -$}}
\put (190,150){\makebox(0,0){$\scriptscriptstyle -$}}
\put (190, 90){\makebox(0,0){$\scriptscriptstyle -$}}
\put (220,120){\makebox(0,0){$\scriptscriptstyle +$}}
\put (250,150){\makebox(0,0){$\scriptscriptstyle +$}}
\put ( 28,138){\makebox(0,0){$\scriptstyle 4$}}
\put ( 88,138){\makebox(0,0){$\scriptstyle 3$}}
\put (148,138){\makebox(0,0){$\scriptstyle 2$}}
\put (208,138){\makebox(0,0){$\scriptstyle 1$}}
\put ( 58,108){\makebox(0,0){$\scriptstyle 3$}}
\put (118,108){\makebox(0,0){$\scriptstyle 2$}}
\put (178,108){\makebox(0,0){$\scriptstyle 1$}}
\put ( 88, 78){\makebox(0,0){$\scriptstyle 2$}}
\put (148, 78){\makebox(0,0){$\scriptstyle 1$}}
\put (118, 48){\makebox(0,0){$\scriptstyle 1$}}
\put ( 58,132){\makebox(0,0){$\scriptstyle 5$}}
\put (118,132){\makebox(0,0){$\scriptstyle 4$}}
\put (178,132){\makebox(0,0){$\scriptstyle 3$}}
\put (238,132){\makebox(0,0){$\scriptstyle 2$}}
\put ( 88,102){\makebox(0,0){$\scriptstyle 5$}}
\put (148,102){\makebox(0,0){$\scriptstyle 4$}}
\put (208,102){\makebox(0,0){$\scriptstyle 3$}}
\put (118, 72){\makebox(0,0){$\scriptstyle 5$}}
\put (178, 72){\makebox(0,0){$\scriptstyle 4$}}
\put (148, 42){\makebox(0,0){$\scriptstyle 5$}}
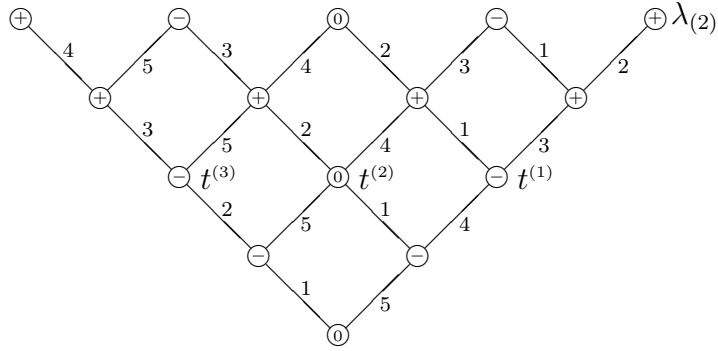
\put (130,10) {\makebox(0,0){Figure 3 : Partition of Weyl orbit of $\la_{(2)}$.
}}
\end{picture}
}

\noindent{\bf A.3 The interaction parameter.}

\noindent It will be shown that the interaction parameter is always positive.
The interaction paramenter is given by
$$A=-{{(\sig_1-\sig_2)^2-(\sig_1 v_1 -\sig_2 v_2)^2-4m^2\sin^2({\pi\over h}(
a_1-a_2))}\over{(\sig_1-\sig_2)^2-(\sig_1 v_1 -\sig_2 v_2)^2-4m^2\sin^2({\pi
\over h}(a_1-a_2))}}$$
\noindent Using the parameterisation $v_{1,2}=\tanh\theta_{1,2}$ (so that for
$-1<v_{1,2}<1$, ${-\infty<\theta_{1,2}<\infty}$) where $\theta_{1,2}$ is the
rapidity of the first and second soliton respectively, and writing $\theta=
\theta_1-\theta_2$ the above can be rewritten
\begin{eqnarray*}
A &=& {{\sin\left({\theta\over 2i}+{\pi (a_p-a_q)\over 2n}\right)\sin\left({
\theta\over 2i}-{\pi (a_p-a_q)\over 2n}\right)}\over{\sin\left({\theta\over
2i}+{\pi (a_p+a_q)\over 2n}\right)\sin\left({\theta\over 2i}-{\pi (a_p+a_q)
\over 2n}\right)}}\cr\cr
&=& {{\cos\left({\pi\over h}(a_p-a_q)\right)-\cos\left({\theta\over i}\right)}
\over{\cos\left({\pi\over h}(a_p+a_q)\right)-\cos\left({\theta\over i}\right)}}
\cr\cr
&=&{{\cos\left({\pi\over h}(a_p-a_q)\right)-\cosh\theta}\over{\cos\left({\pi
\over h}(a_p+a_q)\right)-\cosh\theta}}
\end{eqnarray*}
\noindent As $\theta$ is real, $\cosh\theta\geq 1$, and so the interaction
parameter is always positive.
%
%%%%%%%%%%%%%%%%%%%%%%%%%%%%%%%% REFERENCES %%%%%%%%%%%%%%%%%%%%%%%%%%%%%%%%%%%
%

\end{document}